\documentclass[aps,preprint,prd,showpacs,nofootinbib]{revtex4}

\usepackage{latexsym}
\usepackage{amsmath}
\usepackage{graphicx}
\usepackage{subfigure}
\usepackage{dcolumn}
\usepackage{bm}
\usepackage{amssymb}
\usepackage{latexsym}
\usepackage{ulem}
\usepackage{color}
\usepackage[colorlinks,linkcolor=magenta,anchorcolor=cyan,citecolor=blue]{hyperref}

\def\be{\begin{equation}}
\def\ee{\end{equation}}
\def\ba{\begin{eqnarray}}
\def\ea{\end{eqnarray}}

\def\lf{\left}
\def\rt{\right}

\begin{document}

\title{Oscillation in power spectrum of primordial gravitational wave as a signature
of higher-order stringy corrections}

\author{Yong Cai$^{1}$\footnote{caiyong13@mails.ucas.ac.cn}}
\author{Yu-Tong Wang$^{1}$\footnote{wangyutong12@mails.ucas.ac.cn}}
\author{Yun-Song Piao$^{1,2}$\footnote{yspiao@ucas.ac.cn}}

\affiliation{$^1$ School of Physics, University of Chinese Academy of
Sciences, Beijing 100049, China}

\affiliation{$^2$ State Key Laboratory of Theoretical Physics, Institute of Theoretical Physics, \\
Chinese Academy of Sciences, P.O. Box 2735, Beijing 100190, China}

\begin{abstract}

In low-energy effective string theory, $\alpha'$ corrections
involve the coupling of the dilaton field to Gauss-Bonnet
term. We assume that the dilaton potential is fine tuned so that
the dilaton field may oscillate rapidly for a while around the
minimum of its potential but the inflation background is not
affected. By numerical method, we find that if the dilaton starts
to oscillate at the time of about $\sim 60$ e-folds before the end
of inflation, $\alpha^\prime$ correction may bring unusual
oscillations to the inflationary gravitational wave spectrum,
which might be measurably imprinted in the CMB B-mode
polarization.

\end{abstract}
\maketitle

\section{Introduction}

In the recent years, high-precision CMB observations
\cite{Ade:2015lrj} confirmed the predictions of inflation.
Meanwhile, the searching for the primordial tensor perturbations,
i.e., the primordial gravitational waves (GWs)
\cite{Starobinsky:1979ty},\cite{Rubakov:1982}, which may be
imprinted in the CMB B-mode polarization spectrum
\cite{Kamionkowski:1996zd},\cite{Seljak:1996gy}, has been still on
road, e.g., \cite{Ade:2015tva}. The detection of primordial GWs
would verify general relativity (GR) and strengthen our confidence
in inflation. Inflation occurred in the early universe with higher
scale. Thus, it could be expected that the primordial GWs might
encode information  of the UV-complete quantum gravity
(e.g., \cite{Silverstein:2013wua}), for which the most promising
candidate is string theory.

The effective action of string theory reduces to GR at low
energy. However, it will acquire the higher-order curvature
corrections with the $\alpha'$ expansion. In heterotic string
theory, the leading-order correction is \cite{Metsaev:1987zx} \be
\sim c_1 \alpha'e^{-\phi}R_{GB}^2, \label{RGB}\ee where
$R_{GB}^2$
is the Gauss-Bonnet (GB) term, $M_s=1/\sqrt{\alpha'}$ is the string
scale and $\phi$ is the dilaton field. In Type II string theory,
it is $\sim \alpha^{'3}R^4$, e.g., \cite{Gross:1986iv}, see also
Ref.\cite{Ciupke:2015msa} and references therein.

The effects of $\alpha'$ corrections are actually negligible at
low energy $\ll M_s$. Thus the common expectation is that these
characters of string theory can be hardly directly tested in
laboratory experiments. However, inflation, which occurred in
higher energy scale, is a natural laboratory  for testing
higher energy physics, since the corresponding physics might have
an influence on the primordial perturbations produced during
inflation. The model buildings of inflation in string theory has
been widely investigated, e.g., \cite{Maeda:2004vm} and
\cite{Burgess:2013sla},\cite{Baumann:2014nda} for recent reviews.

The inflationary GWs is a powerful tool to probe the physics
beyond GR, since it is only affected by the physics relevant to
gravity, e.g., the change of the propagating speed of GWs
\cite{Cai:2015dta}. Though the primordial scalar perturbation is
affected also by the modified gravity, it is mainly affected by
the dynamics of inflaton field itself, which will interfere the
identifying of corresponding signatures. In Refs.
\cite{Maldacena:2011nz},\cite{Gao:2011cwa},\cite{Feng:2013pba},\cite{Feng:2014tka},
the non-Gaussianities of inflationary GWs from modified gravity
were studied. However, such non-Gaussianities are far below the
sensitivity of on-road measurements. Comparatively, the power
spectrum of primordial GWs is observationlly promising. Thus a
significant asking is if $\alpha'$ corrections in string theory
may leave measurable imprints in the CMB through their effects on
the power spectrum of GWs.

Here, we assume that the dilaton potential is fine tuned so
that the dilaton field may oscillate rapidly for a while around
the minimum of its potential but the inflation background is not
affected. By numerical method, we find that if the dilaton starts
to oscillate at the time of about $\sim 60$ e-folds before the end
of inflation, $\alpha'$ correction (\ref{RGB}) can lead to the
unusual oscillations in the power spectrum of primordial GWs.
These oscillations may result in some obvious wiggles in the CMB
B-mode polarization spectrum, which is well within the detection
of the upcoming experiments. The intensity of the wiggles is
determined by the string scale $M_s$ and the string coupling
$g_s$.

\section{The setup}

The effective action of heterotic string may be
\cite{Metsaev:1987zx},
\ba S & = & {1\over 2\kappa_4^2}\int  d^4x \sqrt{-g}
\left(R-\frac{1}{2}\partial _{\mu}\phi\partial^{\mu}\phi+ c_1
\alpha'e^{-\phi}R_{GB}^2-V\right)+S_{inf}, \label{L1}\ea where
$\kappa_4^2=1/M_p^2$, and $c_1=1/8$ is dimensionless. The
background is the inflation with $\epsilon=-\dot{H}/H^2\ll
1$, and the field driving inflation is given in $S_{inf}$ (see,
e.g., \cite{Baumann:2014nda}), which we will not involve. The
dimensionless scalar $\phi$ is the dilaton field.  By
variation with respect to metric, we obtain the modified Friedmann equations
\be 6H^2=2{\rho_{inf}\over M_p^2}+{1\over
2}\dot{\phi}^2+V(\phi)+24c_1\alpha^{\prime}e^{-\phi}H^3\dot{\phi}\,,\label{background1}
\ee
\be {\ddot{a}\over a}=-{1\over6M_p^2}(\rho_{inf}+3p_{inf})-{1\over 6}\dot{\phi}^2+{1\over6}V(\phi)+4c_1\alpha^{\prime}e^{-\phi}H^3\dot{\phi}
\lf({1\over2}-\epsilon-{\dot{\phi}\over2H}+{\ddot{\phi}\over H\dot{\phi}}\rt)\,,\label{background2}
\ee
where $\rho_{inf}$ and $p_{inf}$ are the energy density and pressure contributed by inflaton respectively.
Here, for simplicity, we set $\phi$ as a `non-background' field,
i.e. spectator field, which means that it dose not dominate or
affect the evolution of the background. This requires
$\dot{\phi}^2 M_p^2, V M_p^2\ll \rho_{inf}\simeq H^2 M_p^2$, and
the contribution of $c_1\alpha' e^{-\phi}{\dot \phi}H^3$ from the
GB term to $H^2$ must be also negligible, which put the constraint
$c_1\alpha' e^{-\phi}{\dot \phi}H\ll 1$.
In addition, \be {H^2\over M_s^2g_s}\ll 1 \label{MSGS}\ee
must be imposed, so that other higher-order $\alpha'$
corrections may be neglected, where $g_s=e^{\phi}$ is the string coupling.

The dilaton $\phi$ is massless in the weak coupling regime, i.e.,
$\phi\ll -1$ and $g_s\ll 1$. However, in the moderately strong
coupling regime, $\phi\sim -1$, the potential must exhibit some
structure to make $\phi$ trapped at the minimum of the potential,
so that the equivalence principle is preserved at late time
\cite{Damour:2002mi}. Though the detail of the dilaton potential
$V(\phi)$ is still an issue in the study, it is generally thought
that $V(\phi)$ with a local valley can naturally arise due to some
nonperturbative effects \cite{Silverstein:2004id}.

Inflation happens when inflaton dominates over other fields.
We can see from Eq.(\ref{background1}) that it's unlikely
$\phi\ll-1$ at the beginning of inflation, since the term with
$e^{-\phi}$ will dominate over the contribution from inflaton.
Thus, it's natural to assume that $\phi$ locates at the strong or
moderately strong coupling regime with $\phi\simeq -1$ at
the beginning of inflation, and runs towards the regime
with $\phi<-1$ until trapped by the valley of $V(\phi)$.
In addition, we also require that the initial value of
$V(\phi)$ must also be negligible compared with the inflaton
potential.

In this setup, the adiabatic scalar perturbation is
contributed by inflaton, which is irrelevant with the
`non-background' field $\phi$.
We, following \cite{Maldacena:2002vr}, may obtain the quadratic
action of the GWs mode $h_{ij}$ as follows
\ba S^{(2)}=M_p^2 \int
d\tau d^3x\, {a^2 Q_T\over 8} \left[ h_{ij}'^2-c_T^2(\vec{\nabla}
h_{ij})^2 \right],\label{action2}
\ea
where $h_{ij}$ satisfies
$\partial _i h_{ij}=0$ and $h_{ii}=0$, $\tau$ is the conformal
time, $d\tau=dt/a $, a prime denotes the derivative with respect
to $\tau$, $c_T$ is the propagating speed of primordial GWs, and
$Q_T M_p^2$ is regarded as effective Planck scale
$M_{P,eff}^2(\tau)$, and
\be
c_T^2=\frac{1}{Q_T}\left(4c_1\alpha'{\ddot \xi}+1\right),\,\,\,\,
Q_T=4c_1\alpha' {\dot \xi}H+1,\label{ct}
\ee
where
$\xi=1/e^{\phi}$, which are consistent with the earlier results in
Ref.\cite{Kawai:1998ab}\cite{Cartier:2001is}. Thus, although
$\phi$ dose not dominate the background, due to its coupling to
$R_{GB}^2$, it may affect the tensor perturbation, see also
\cite{Feng:2013pba} for similar case but with nonminimal
derivative coupling to $R$. Here, the conditions of
avoiding the ghost instability are $c_T^2>0$ and $Q_T>0$.

Here, if $\phi$ sits in the minimum of its potential or slow
rolls, we have ${\dot{Q_T} }/{HQ_T}\ll1$ and $ {\dot{c_T} }/{H
c_T}\ll 1$, the primordial GWs spectrum will only acquire a small
correction \cite{Jiang:2013gza},\cite{DeFelice:2014bma}, and its
shape will not be altered essentially. However, before
resting at the minimum of its potential, if $V_{\phi\phi}\gg H^2$,
the dilaton field might rapidly roll down towards the minimum of
its potential and oscillate around it, which will lead to a
short-time but drastic variation of the propagating speed of GWs,
see (\ref{ct}). We will show that it is this variation, if
happened during inflation, that will naturally imprint the
leading-order string correction (\ref{RGB}) to the
large-scale primordial GWs.

In the pre-big bang scenario, the $\alpha'$ corrections
were applied to solve the singularity problem of the big
bang model, which will also leave imprints in the primordial GWs
\cite{Gasperini:1997up},\cite{Cartier:2001gc}. However, these
imprints appear at small scales, and are not relevant to the CMB
polarization, unless the pre-big bang phase is followed by an
inflationary phase \cite{Piao:2003zm},\cite{Liu:2013kea}, or see \cite{Satoh:2007gn}.

\section{The signature of higher-order string correction}

The power spectrum of primordial GWs is calculated as follows. The
Fourier series of $h_{ij}$ is \be h_{ij}(\tau,\mathbf{x})=\int
\frac{d^3k}{(2\pi)^{3} }e^{-i\mathbf{k}\cdot \mathbf{x}}
\sum_{\lambda=+,\times} \hat{h}_{\lambda}(\tau,\mathbf{k})
\epsilon^{(\lambda)}_{ij}(\mathbf{k}),\ee where $
\hat{h}_{\lambda}(\tau,\mathbf{k})=
h_{\lambda}(\tau,k)a_{\lambda}(\mathbf{k})
+h_{\lambda}^*(\tau,-k)a_{\lambda}^{\dag}(-\mathbf{k})$, the
polarization tensors $\epsilon_{ij}^{(\lambda)}(\mathbf{k})$
satisfy $k_{j}\epsilon_{ij}^{(\lambda)}(\mathbf{k})=0$,
$\epsilon_{ii}^{(\lambda)}(\mathbf{k})=0$,
 and $\epsilon_{ij}^{(\lambda)}(\mathbf{k})
\epsilon_{ij}^{*(\lambda^{\prime}) }(\mathbf{k})=\delta_{\lambda
\lambda^{\prime} }$, $\epsilon_{ij}^{*(\lambda)
}(\mathbf{k})=\epsilon_{ij}^{(\lambda) }(-\mathbf{k})$, the
commutation relation for the annihilation and creation operators
$a_{\lambda}(\mathbf{k})$ and
$a^{\dag}_{\lambda}(\mathbf{k}^{\prime})$ is $[
a_{\lambda}(\mathbf{k}),a_{\lambda^{\prime}}^{\dag}(\mathbf{k}^{\prime})
]=\delta_{\lambda\lambda^{\prime}}\delta^{(3)}(\mathbf{k}-\mathbf{k}^{\prime})$.
Thus we have the equation of motion for $u(\tau,k)$ as
\be
u''+\left(c_T^2k^2-\frac{z_T''}{z_T} \right)u=0, \label{eom1}
\ee
where
\be {u}(\tau,k)= h_{\lambda}(\tau,k)\cdot {z_T}, \quad z_T=
{aM_p { \sqrt{Q_T} }\over 2}.\label{zt}
\ee
Initially, the
perturbations are deep inside the horizon, i.e., $c_T^2k^2 \gg
\frac{z_T''}{z_T}$, the initial condition is $u\sim
\frac{1}{\sqrt{2c_T k} }e^{-i c_T k\tau}$. The power spectrum of
GWs is
\be
P_T=\frac{k^3}{2\pi^2}\sum_{\lambda=+,\times}
\lf|h_{\lambda} \rt|^2=\frac{4k^3}{\pi^2 M_p^2}\cdot\frac{1}{Q_T
a^2} \lf|u \rt|^2, \quad aH/k \gg 1.\label{pt}
\ee
Here, since
both $c_T$ and $Q_T$ experience the short-time but drastic
variations, analytical solution is difficult to obtain. We will
first exhibit the numerical results, which help to clearly
display the nontrivial feature of GWs spectrum induced by
$\alpha'$ correction.

\subsection{The numerical solution}

To facilitate the numerically solving of the perturbation
equation, we introduce the variable $\alpha=\ln a$
\cite{Adams:2001vc}. Using
$\frac{d}{dt}=H(\alpha)\frac{d}{d\alpha}$, the evolving equation
of $\phi$ and Eq. (\ref{eom1}) can be rewritten as
\be
\phi_{\alpha\alpha}+\lf(3+\frac{H_{\alpha}
}{H}\rt)\phi_{\alpha}+24\lf(1+\frac{H_{\alpha} }{H} \rt)
H^2c_1\alpha' e^{-\phi}  +\frac{V_{\phi} }{H^2}=0, \label{eom}
\ee
\be u_{\alpha\alpha}+\lf(1+\frac{ H_{\alpha} }{H}
\rt)u_{\alpha}+\frac{1}{a^2 H^2
}\cdot\lf(c_T^2k^2-\frac{z_T''}{z_T} \rt)u=0, \label{eom2} \ee
where the subscript `$\alpha$' denotes ${\partial/\partial
\alpha}$, and $c_T^2$ and $z_T$ are given by Eqs. (\ref{ct}) and
(\ref{zt}), respectively. The initial condition of Eq.
(\ref{eom2}) is written as   \be u|_{\alpha \ll
\ln(\frac{k}{H})}=\frac{1}{\sqrt{2k} }, \quad \frac{\partial u}{\partial
\alpha}\Big|_{\alpha \ll \ln(\frac{k}{H})}=-i \sqrt{\frac{k}{2}
}\frac{1}{e^{\alpha} H}\Big|_{\alpha \ll \ln(\frac{k}{H})}, \ee where the
initial value of $c_T$ is set as unity. We assume, without
loss of generality, that the background of slow-roll inflation is
described by constant $\epsilon$, so that the Hubble parameter may be
written as $H(\alpha)/M_p=A_H\cdot e^{-\epsilon \cdot \alpha}$, in
which $A_H$ is constant.

As has been argued, in the moderately strong coupling regime
$\phi\sim -1$, due to the non-perturbative effects, the potential
must exhibit some structure to make $\phi$ trapped at the minimum
of the potential. Initially, $\phi$ might deviate from the minimum
of its potential. In (\ref{L1}), the dimension of $V(\phi)$ is the
square of mass and $\phi$ is dimensionless. In term of Ref.
\cite{Silverstein:2004id}, both $V_\phi$ and $V_{\phi\phi}$ are
$\sim M_p^2e^{2\phi}l_s^6/{\cal V}_6=M_s^2$, in which ${\cal V}_6$
is the volume of extra dimensions. Thus with (\ref{MSGS}) and
$g_s=e^{\phi_*}\ll 1$, we have $V_{\phi\phi}\gg H^2$. Therefore,
$\phi$ will rapidly roll down, and oscillate around the minimum of
the potential before getting rest. As long as $V_\phi\sim
V_{\phi\phi}\sim M_s^2$ are fixed (note that $V(\phi)\ll 6H^2$ is
still reserved), the result is actually insensible to other
details of $V(\phi)$.

Here, for the purpose of numerical simulation, we set the
potential locally as $V(\phi)\sim {1\over 4}V_0(\phi-\phi_*)^4 $,
where $\phi_*<-1$. To not interrupt inflation, it's convenient to
parameterise the potential locally as \be V(\phi)=V_0\cdot
\frac{(\phi-\phi_*)^4}{1+B\cdot M_p^{-2}V_0
(\phi-\phi_*)^4},\label{Vpara} \ee where $V_0$ has the dimension
of the square of mass, $B$ is a dimensionless constant. The
purpose of introducing $B$ is to guarantee that $V(\phi)$
doesn't affect the background evolution when $|\phi-\phi_*|$ is
large enough, as we can see from Eq.(\ref{background1}) and
(\ref{background2}). We will set $B=10^{10}$, then the value of
$V(\phi)$ is smaller than $10^{-10}M_p^2$, thus is negligible
compared with $6H^2$. There is some finetunning in the
shape of the potential. The shape of dilaton potential is
actually still an issue in study, and it is generally thought that
in heterotic string theory the potential with a local valley might
arise due to some nonperturbative effects. What we mainly focused
on is that how the primordial GWs
spectrum will be affected, if there is such a potential.

Because both the contributions from the Gauss-Bonnet and the potential is
to move $\phi$ towards smaller value, $\phi$
won't be frozen if it located at $|\phi-\phi_*|>1$ initially, and
we don't need to create very fine-tuned initial conditions for
$\phi$ \footnote{ We plot $\phi(\alpha)$ with different initial
value $\phi_{initial}$ in Fig.\ref{diff-phi},}. Below, we will
set $\phi_*=-2.6$, and the initial value of $\phi$ as
$\phi_{initial}=-1.5$.

\begin{figure}[htbp]
\subfigure[~~$\phi(\alpha)$, $\phi'(\alpha)$ and $\phi''(\alpha)$]{\includegraphics[width=.47\textwidth]{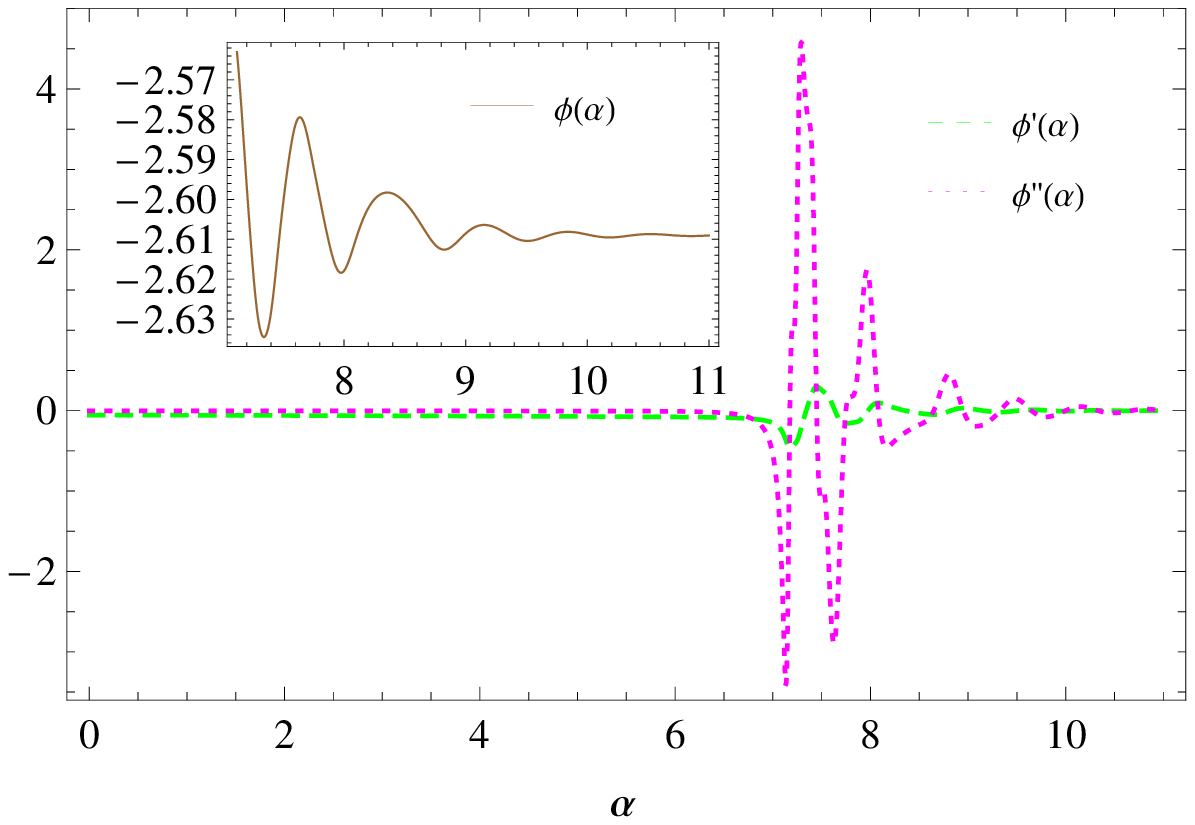} }
\subfigure[~~$c_T^2$ and
$Q_T$]{\includegraphics[width=.48\textwidth]{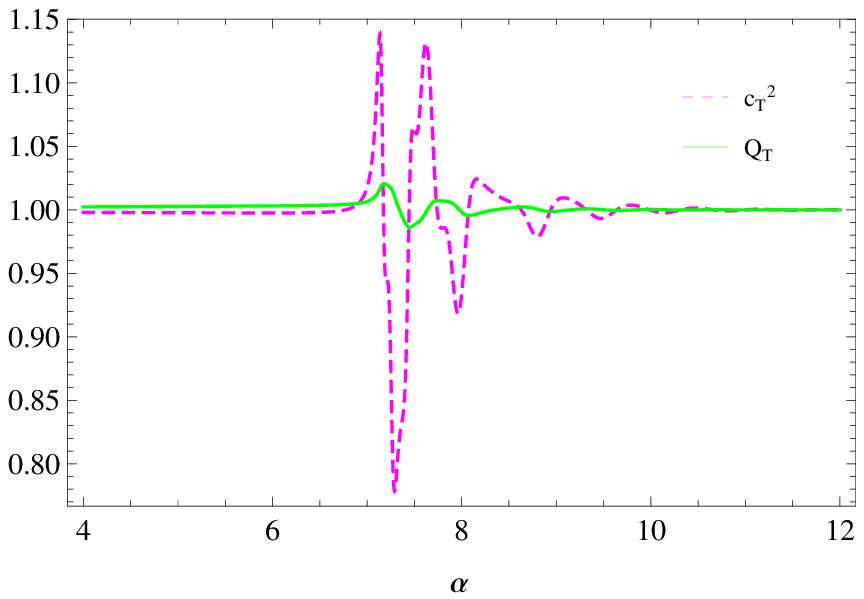} }
\caption{The evolutions of the dilaton field $\phi$, $c_T^2$ and
$Q_T$ for $\epsilon=0.003$, $A_H=2.72\times10^{-5}$, $c_1=1/8$,
$\alpha'M_p^2=9.6\times10^6$, $V_0/M_p^2=6.5\times 10^{-5}$,
$B=10^{10}$, $\phi_{*}=-2.6$, $\alpha_{initial}=-13.4$ and $\phi_{initial}=-1.5$.} \label{fig01}
\end{figure}

In Fig.\ref{fig01}, we plot the evolutions of $\phi$, $c_T^2$ and
$Q_T$.  The field $\phi$ rolls down along its potential, and gets
rest at the minimum of the potential after oscillating around it,
since the inflation rapidly dilutes its kinetic energy. This
evolution induces the oscillations of $c_T^2$ and $Q_T$.
The oscillation amplitude of $c_T^2$ is far larger than that of
$Q_T$, since $c_T^2$ is dominated by $\ddot \phi$ while $Q_T$ is
by $\dot \phi$, and $\dot\phi$ can hardly be large enough since
${\dot \phi}^2 M_p^2\ll \rho_{inf}\simeq H^2M_p^2$.
We can see that both $c_T^2>0$ and $Q_T>0$ are always satisfied in the regime we are interested in, thus there is no ghost instability.
Here,
from the values of relevant parameters, we have
$H^2/(M_s^2g_s)\simeq 0.1$, which satisfies the condition
(\ref{MSGS}).

\begin{figure}[htbp]
\includegraphics[scale=2,width=0.55\textwidth]{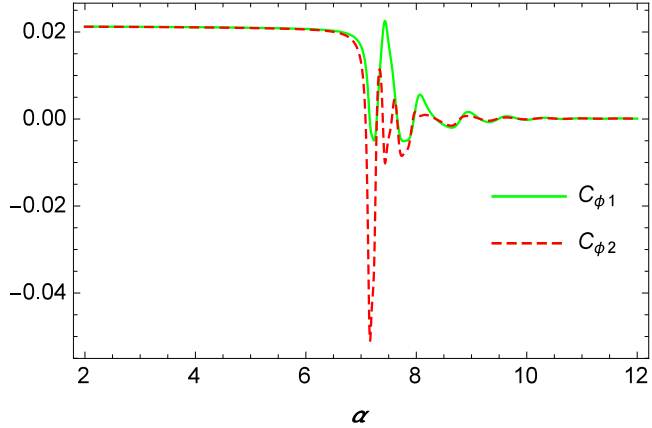}
\caption{The contributions from non-trivial variation of $\phi$ and its derivatives to the background evolution, where $\epsilon=0.003$, $A_H=2.72\times10^{-5}$, $c_1=1/8$,
$\alpha'M_p^2=9.6\times10^6$, $V_0/M_p^2=6.5\times 10^{-5}$, $B=10^{10}$,
{$\phi_{*}=-2.6$, $\alpha_{initial}=-13.4$ and $\phi_{initial}=-1.5$.} We can see that $C_{\phi}\ll1$ in the regime we are interested in.} \label{figconstraint}
\end{figure}

It might be concerned that whether the contribution of the
spectator scalar field $\phi$ to the background evolution can be
safely negligible. To verify this issue, we set the contribution
of $\phi$ in Eq.(\ref{background1}) and (\ref{background2}) as \be
C_{\phi_1}=\lf({1\over
2}\dot{\phi}^2+V(\phi)+24c_1\alpha^{\prime}e^{-\phi}H^3\dot{\phi}\rt)\Big/(6H^2),
\label{Cphi1}\ee \be C_{\phi2}=\lf[-{1\over
6}\dot{\phi}^2+{1\over6}V(\phi)+4c_1\alpha^{\prime}e^{-\phi}H^3\dot{\phi}
\lf({1\over2}-\epsilon-{\dot{\phi}\over2H}+{\ddot{\phi}\over
H\dot{\phi}}\rt) \rt]\Big/\lf({\ddot{a}\over a}\rt), \ee
respectively. We plot $C_{\phi_1}$ and $C_{\phi_2}$ with respect
to $\alpha$ in Fig.\ref{figconstraint}, using the same parameters
used in Fig.\ref{fig01}. We can see that the contributions from
the terms relevant with $\phi$ are negligible compared with that
of inflaton in the regime we are interested in.

\begin{figure}[htbp]
\includegraphics[scale=2,width=0.55\textwidth]{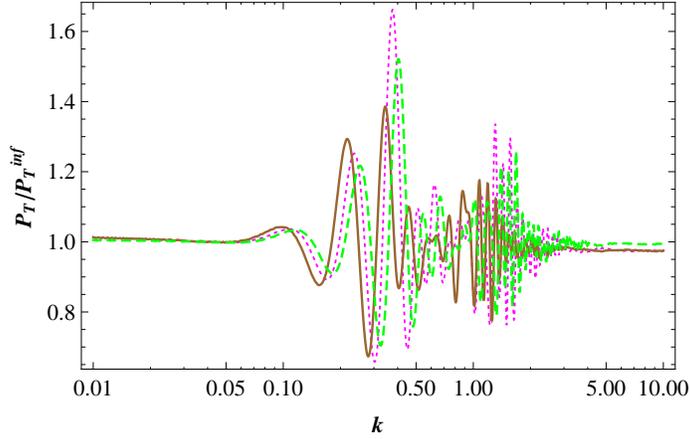}
\caption{Power spectra of primordial GWs, in which
$A_H=2.72\times10^{-5}$, $c_1=1/8$, $\phi_{*}=-2.6$, $\alpha_{initial}=-13.4$ and $\phi_{initial}=-1.5$, while
$\epsilon=(0.003,\,0.001,\,0.003 )$,
$V_0/M_p^2=(6.5,\,12,\,12)\times 10^{-5}$, $B=10^{10}$  and
$\alpha'M_p^2=(9.6\times10^6,\,8.0\times10^6,\,9.6\times10^6 )$,
for brown solid curve, green dashed curve, and magenta dotted
curve, respectively.} \label{fig02}
\end{figure}

\begin{figure}[htbp]
\subfigure[~~$r=16\epsilon=0.016$, $V_0/M_p^2=6.5\times
10^{-5}$]{\includegraphics[width=.48\textwidth]{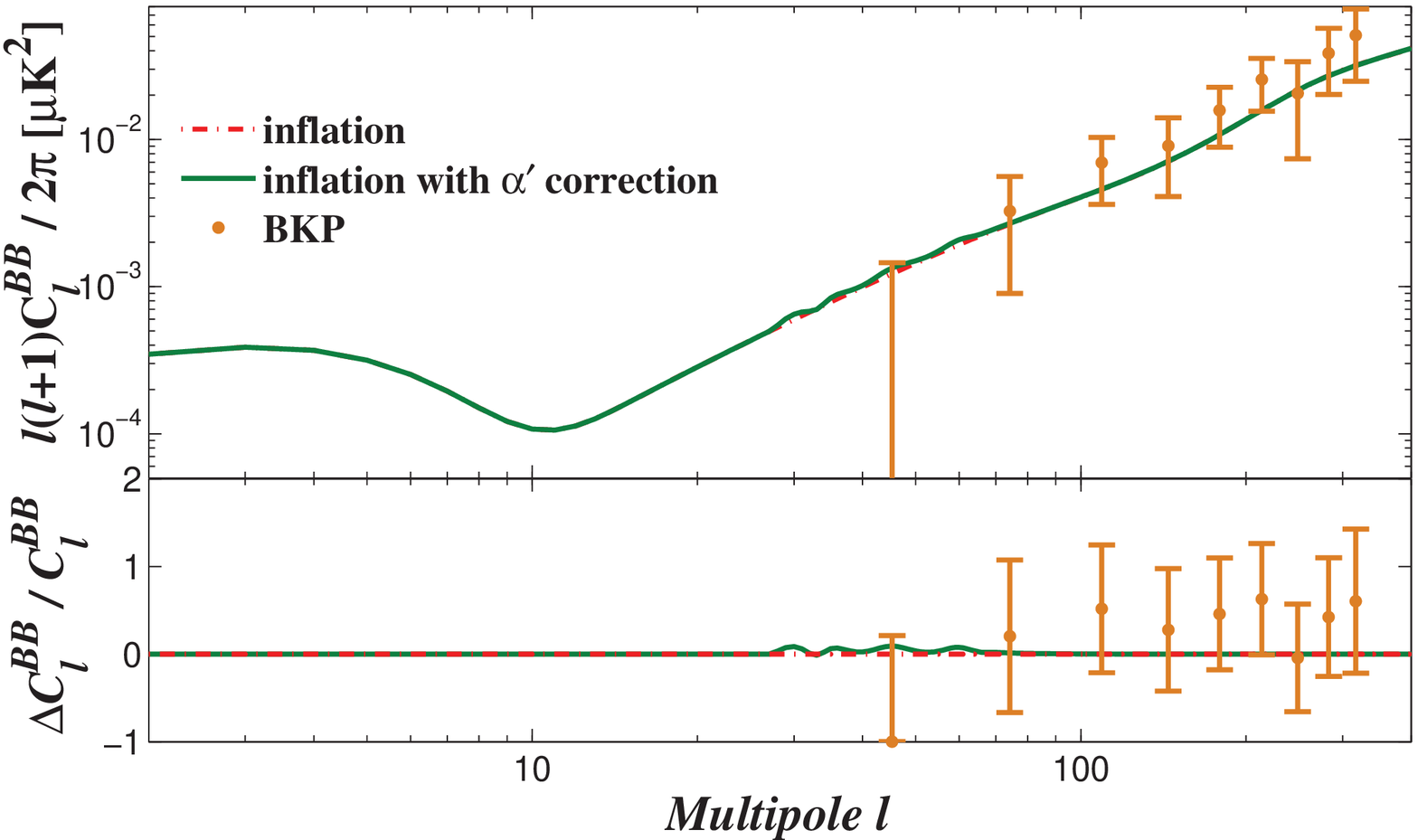}
} \subfigure[~~$r=16\epsilon=0.048$, $V_0/M_p^2=1.2\times
10^{-4}$]{\includegraphics[width=.47\textwidth]{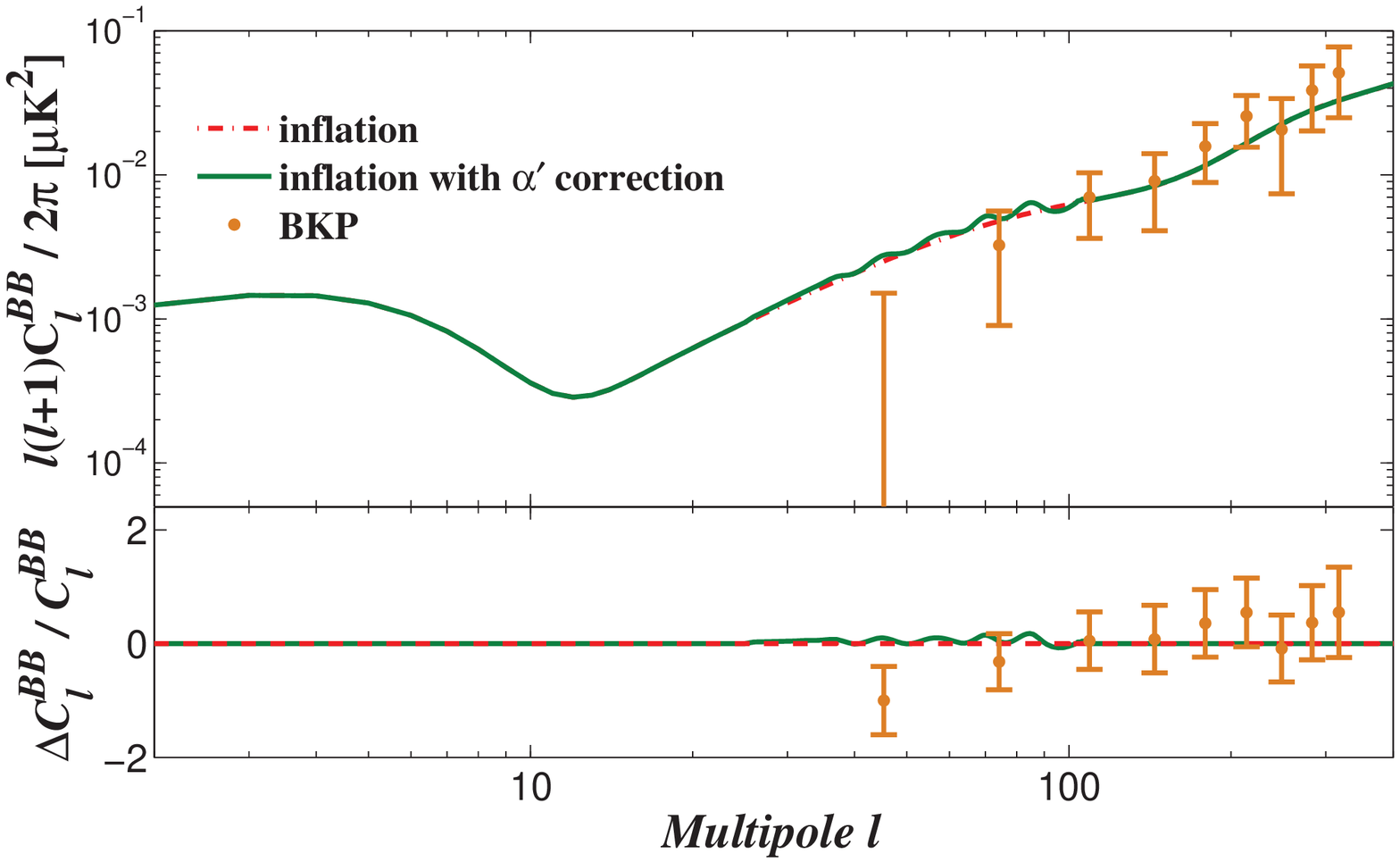}
} \caption{Wiggle features in the CMB lensed B-mode power spectra,
in which $A_H=2.72\times10^{-5}$, $c_1=1/8$,
$\alpha'M_p^2=9.6\times10^6$, $B=10^{10}$, $\phi_{*}=-2.6$, $\alpha_{initial}=-13.4$, $\phi_{initial}=-1.5$, and
$r=P_T^{inf}/P_{\cal R}^{inf}$. Both $P_T^{inf}$ and $P_{\cal
R}^{inf}$ are those of slow-roll inflation without $\alpha'$
correction. BKP data are from the joint analysis of
BICEP2/KeckArray and Planck \cite{Ade:2015tva}. } \label{fig03}
\end{figure}

In Fig.\ref{fig02}, we plot the ratio of the power spectrum $P_T$
to $P_T^{inf}$ for different $\alpha'$, $V_0$ and $\epsilon$, in
which $P_T^{inf}={2H^2\over \pi^2 M_p^2}$ is that of standard
inflation without $\alpha'$ correction. In all cases, unusual
oscillations in $P_T$ are present, which is obviously a universal
character attributed to the nontrivial behavior of $c_T$ and
$Q_T$, or more intrinsically said, the string theory might
manifest. Though the particle production during inflation may also
significantly affect the power spectrum of GWs
\cite{Cook:2011sp},\cite{Senatore:2011sp}, it only leads to a
bumplike modification, which is entirely different from the
behavior of the oscillation showed here.

In Fig.\ref{fig03}, we plot the CMB lensed BB-mode power
spectrum, compared with that of inflation without $\alpha'$
correction, in which $r=P_T^{inf}/P_{\cal R}^{inf}$, and
both $P_T^{inf}$ and $P_{\cal R}^{inf}$ are those of standard
inflation. Here, we have assumed that the power spectrum of
the scalar perturbation is not affected by the nontrivial
behavior of the dilaton $\phi$, and is still $P_{\cal R}^{inf}$.
In addition, we also assumed that after inflation the dilaton
rests at the minimum of its potential so that GR is covered. Thus,
the effect of varying $c_T$ at late time may be neglected, or see
\cite{Amendola:2014wma} \cite{Raveri:2014eea}. When the dilaton field $\phi$ is fixed,
the GB term will not affect the evolution of the background,
since it is the generalised Euler density of four-dimensional
spacetime, and thus is a topological invariant and does not
contribute to the equations of motion.

There are generally a recombination peak at $l\approx 80$ and a
reionization bump at $l < 10$. Both are the imprints of the
primordial GWs. The detection of the recombination peak will be a
confirmation that inflation has ever occurred, while
detecting the reionization bump at low-$l$ would help us to
understand the physics of pre-inflationary universe,
e.g.,\cite{Cai:2015nya}. When the comoving oscillating scale in
GWs power spectrum is {set at} $l\approx 80$, for $r\gtrsim 0.01$
we may see obvious wiggles around the recombination peak, which
may be tested by the upcoming polarization data.

\subsection{The analytic estimate}

The oscillations in $P_T$ are induced by $\alpha'$
correction, which naturally implies that the intensity of the
oscillations is closely related to the value of the string scale
$M_s=1/\sqrt{\alpha'}$. We have observed in
Fig.\ref{fig02} that with larger $\alpha'$, i.e., smaller energy
scale of string $M_s$, we got larger amplitude of the oscillation
in the power spectrum. To acquire more insights on relevant
physics, we will give an analytic estimate for it.

To not affect the background evolution, $\phi_{\alpha}^2\ll 12$,
$V\ll 6H^2$ and $4H^2 c_1\alpha' e^{-\phi}\phi_{\alpha}\ll 1$
must be satisfied. Thus we have $Q_T\approx 1$ and $c_T^2\approx
1-4H^2 c_1\alpha'e^{-\phi}\phi_{\alpha\alpha}$. In addition, we
have $V_\phi/H^2\simeq M_s^2/H^2\gg 1$, so when $\phi$ rapidly
rolls down and oscillates around the minimum of its potential,
which corresponds to $\phi=\phi_*$, in Eq. (\ref{eom})
$V_{\phi}/H^2$ dominates $\phi_{\alpha\alpha}$, i.e.,
$\phi_{\alpha\alpha}\approx V_{\phi}/H^2$. The second term in Eq.
(\ref{eom}) is not negligible, which is responsible for making the
oscillation stop rapidly. Thus our estimate is precise only
for the first two or three oscillations. However, it is
enough for us to grasp the physics behind the oscillating $P_T$.
When $\phi$ oscillates around $\phi_*$, $c_T^2$ will acquire an
oscillation with same frequency. The extreme values of the
oscillating amplitude of $c_T^2$ can be estimated as \ba c^2
_{T\,max(min)}\approx 1\pm4 c_1\alpha' e^{-\phi_*}|V_{\phi}|,
\label{cTestimation} \ea where $e^{-\phi}$ is approximately
replaced with $e^{-\phi_*}$. Therefore, for $V_{\phi}\sim M_s^2$,
we may have $c_1\alpha' e^{-\phi_*}|V_{\phi}|\gtrsim 0.1$, which
is required for the spectrum showing itself feature.

We assume that $c_T$ experiences an ideally steplike jump from
$c_{T1}$ to $c_{T2}$, and following
Ref. \cite{Cai:2015dta} (see, also \cite{Nakashima:2010sa} for the
scalar perturbation),  we have \be
P_T=P_T^{inf}\cdot\frac{f(k,k_0,x)}{Q_Tc^3_{T1}}, \ee where
\be f(k,k_0,x)\approx \frac{1}{x^2}\left[1+({1\over
x^2}-1)\cos^2(\frac{k}{k_0}) \right] \ee for $k\gg k_0$, which
oscillates between $1/x^2$ and $1/x^4$, and $k_0$ is the wave
number of the GWs mode crossing horizon when $c_T$ jumps, and
$x={c_{T2}\over c_{T1}}$.

Thus we approximately have \ba \lf(\frac{P_T}{P_T^{inf} }
\rt)_{max}\approx \frac{c_{T\,max} }{c^4_{T\,min} }\, ,\quad
\lf(\frac{P_T}{P_T^{inf} } \rt)_{min}\approx \frac{c_{T\,min}
}{c^4_{T\,max} }\, .\label{PTestimation} \ea
This reflects the effect of $\alpha'$, which corresponds to $M_s$,
on the amplitude of oscillation in the power spectrum. In turn, we
have \ba {M_p^2\over M_s^2}=\alpha' M_p^2 \approx \frac{g_s
}{4c_1|{\tilde V}_{\phi}| }\cdot
\frac{\lf[\frac{(P_T/P_T^{inf})_{max} }{(P_T/P_T^{inf})_{min} }
\rt]^{2/5}-1 }{\lf[\frac{(P_T/P_T^{inf})_{max}
}{(P_T/P_T^{inf})_{min} } \rt]^{2/5}+1}\, . \label{energy} \ea
where $g_s=e^{\phi_*}$ and ${\tilde V}_{\phi}=V_{\phi}/M_p^2\ll 1$
is dimensionless. This indicates that if the upcoming CMB
B-mode polarization experiments would uncover wiggles in
the B-mode power spectrum, it may offer important information
about $M_s$, the string coupling $g_s$ and the shape of the
dilaton potential.

\begin{figure}[htbp]
\includegraphics[scale=2,width=0.55\textwidth]{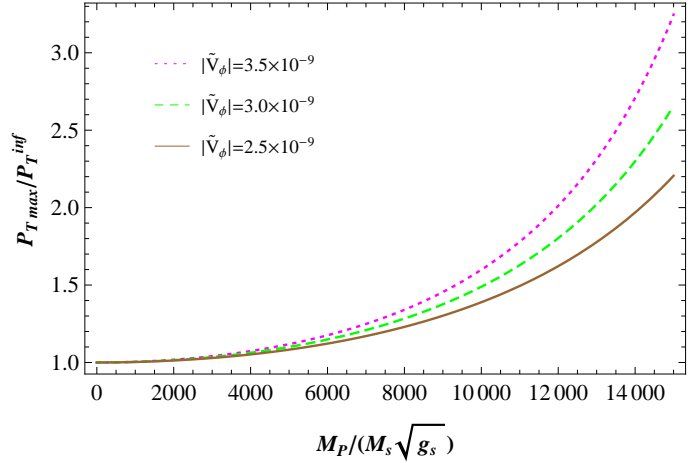}
\caption{The relation between $P_{T\,max}/P_{T}^{inf}$, $M_p/(M_s
\sqrt{g_s})$ and $|\tilde{V}_{\phi}|$. } \label{fig04}
\end{figure}

In Fig.\ref{fig04}, we plot $P_{T\,max}/P_{T}^{inf}$ with respect
to $M_p/(M_s \sqrt{g_s})$ for different values of $|{\tilde
V}_{\phi}|$, in which $c_1=1/8$ for the heterotic string. To keep
$c_T^2>0$, we require $M_p/(M_s \sqrt{g_s})\lesssim
\sqrt{2/|\tilde{V}_{\phi}| }$. In addition, the result of Planck
\cite{Ade:2015lrj} showed the scalar perturbation amplitude
$P_{\cal R}^{inf}\approx 3\times 10^{-9}$, so (\ref{MSGS})
becomes $M_p/(M_s \sqrt{g_s})\ll |M_p/H|\approx 36755$ for
$r=0.05$. In performing Fig.\ref{fig04}, all above conditions are
satisfied.

The power spectra plotted in Fig.\ref{fig02} indicates
$\alpha'M_p^2 \approx 8\times 10^6$, $5.4\times10^6$ and
$7.4\times10^6$ for the magenta dotted, green dashed and brown
solid curves, respectively, which are consistent with Eq.
(\ref{energy}) and Fig.\ref{fig04}. Thus Eq. (\ref{energy})
actually works well for the first two or three
oscillations. This provides a direct link between the oscillating
amplitude in the primordial GWs spectrum and the string
parameters, which is depicted in Fig.\ref{fig04}.

In heterotic string theory, the spacetime is 10-dimensional. The
leading term of the effective action is $S_{10} ={1\over
2\kappa_{10}^2}\int d^{10}x\sqrt{-G}e^{-2\phi}R_{10}$ in the string
frame, in which $\kappa_{10}^2\sim \alpha'^4$. After compactifying
it on a Calabi-Yau manifold with volume ${\cal V}_6$, in
four-dimension we have $M_p^2={\cal V}_6/(g_s^2\kappa_{10}^2)$,
e.g., \cite{Becker:2007}. Thus combining it with Fig.\ref{fig04},
in which we see $M_p/(M_s \sqrt{g_s})\sim 10^4$ for
$P_{T\,max}/P_T^{inf}\gtrsim 1.1$, we have \be {{\cal V}_6\over
l_s^6}={g_s^2M_p^2\over M_s^2}\sim 10^8g_s^3>1.
\label{V6}\ee for $g_s\sim 0.1$. This is consistent with the
requirement of supergravity approximation, i.e., ${\cal V}_6>
\alpha'^3$.

Here, for simplicity, we required that the background is not
affected by the dilaton field. However, it is also possible that
the dilaton field would affect the background at certain level. In
that case, it is required to evaluate the effect of $\alpha'$
correction on the scalar perturbation spectrum. Additionally, the
oscillation of primordial GWs power spectrum will also affect the
auto-bispectrum of the tensor perturbations and the
cross-bispectrum of the primordial perturbations, which
will be studied in upcoming work.

\section{Conclusion}

The $\alpha^\prime$ correction (\ref{RGB}) can be regarded
as high-order correction to GR. Due to some nonperturbative
effects, the dilaton potential might show itself a local valley.
We assume that the dilaton potential is fine tuned so that the
dilaton field may oscillate rapidly for a while around this
valley but the inflation background is not affected.

We numerically find that if the dilaton starts to oscillate at the
time of about $\sim 60$ e-folds before the end of inflation, the
correction (\ref{RGB}) may bring unusual oscillations to the
inflationary GWs spectrum, which might be measurably imprinted in
the CMB B-mode polarization. We analytically show that the
intensity of oscllating the is determined by the string scale
$M_s$ and the string coupling $g_s$.

We show that if $r\gtrsim 0.01$, we might observe obvious wiggles
around the recombination peak in the CMB B-mode power spectrum,
which may be well within the detection of the current B-mode
polarization experiments. This finding, if supported by
observations, would yield significant insights into the gravity
theory beyond GR. Conversely, a null result would place tight
constraints on corresponding theory. Our work highlights the fact
that high-precision CMB B-modes polarization experiments might
offer us richer information on a UV-complete gravity theory than
expected.

\textbf{Acknowledgments}

We thank Mingzhe Li, Jianxin Lu and Kaixi Feng for valuable discussions. This
work is supported by NSFC, No. 11222546, and National Basic
Research Program of China, No. 2010CB832804, and the Strategic
Priority Research Program of Chinese Academy of Sciences, No.
XDA04000000. We acknowledge the use of CAMB.

\appendix

\section{The isocurvature perturbations from $\phi$}

The curvature perturbation $\cal R$ is induced by inflaton, while
the perturbation of the spectator field $\phi$ is the isocurvature
perturbation. By perturbing $\phi$ with $\delta\phi$, we have
\ba S^{(2)}_{\delta\phi}=\int d^4x {a^3M_p^2\over 4} &
\big[ & \delta\dot{\phi}^2 -
\partial_i\delta\phi\partial^i\delta\phi\nonumber\\ & -&
2c_1\alpha'e^{-\phi} H^2\left(\partial^i
h_{ij}\partial^j\delta\phi+2\partial_i {\cal R}\partial^i
\delta\phi+H {\dot {\cal R}}\delta\phi\right)
 \nonumber\\
& + &
\left(c_1\alpha'e^{-\phi}R_{GB}^2-V_{\phi\phi}\right)(\delta\phi)^2
\big].\label{deltaphi} \ea Thus $\delta\phi$ is instability-free.
Actually, we have $c_1\alpha^\prime e^{-\phi}H^2\ll 1$, since
$c_1\alpha^\prime e^{-\phi}H{\dot\phi}\ll 1$ and ${\dot\phi}\ll
H$, see Eq.(\ref{Cphi1}), otherwise $\phi$ will dominate the
background. Thus during inflation the curvature perturbation $\cal
R$ is hardly affected by the spectator scalar field $\phi$. In
addition, when perturbed, $e^{-\phi}R_{GB}^2$ also contributes
such as \be \sim {\dot {\cal R}}\delta\phi\delta\phi,\,\,\,
\partial_i{\cal R}\partial^i{\cal R}\delta\phi, \ee which are
higher-order and do not appear in the quadratic action of $\cal
R$. This implies that as long as inflaton is canonical, $\cal R$
is also instability-free.

However, after inflation the isocurvature perturbation from $\phi$
possibly converts to the curvature perturbation. In this Appendix,
we will demonstrate that the isocurvature perturbation from $\phi$
is negligible.

By defining $v=a\delta\phi$, the perturbation equation of
Eq.(\ref{eom}) is given by
\be {d^2v\over d\tau^2}+\lf[k^2-{1\over
\tau^2}\lf(\nu^2-{1\over4}\rt) \rt]v=0, \label{vv}\ee where
$24c_1\alpha'e^{-\phi}H^4(1-\epsilon)\ll V_{\phi\phi}$ has been
neglected and $\nu^2= {9\over 4}-{V_{\phi\phi}\over H^2}$.

We can set $V_{\phi\phi}= {m_{\phi}^2}$ constant for an analytic
estimate \cite{Riotto:2002yw}.
Then the solution can be given as
\be v={\sqrt{\pi}\over2}e^{i\lf(\nu+{1\over2}\rt){\pi\over 2}}\sqrt{-\tau}H_{\nu}^{(1)}(-k\tau),\label{v}
\ee
When $m_{\phi}\ll H$, the power
spectrum of $\delta\phi$ is nearly scale-invariant. However, when
$m_{\phi}\gg H$, $\nu$ is imaginary. For convenience, we set $\hat{\nu}=-i\nu$. Using that $H_{i\hat{\nu}}^{(1)}(x\ll1)\sim (x/2)^{i\hat{\nu} }/\Gamma(1+i\hat{\nu})-i(x/2)^{i\hat{\nu} }\cosh(\pi\hat{\nu})\Gamma(-i\hat{\nu})/\pi-i(x/2)^{-i\hat{\nu} }\Gamma(i\hat{\nu})/\pi$, we have $|H_{i\hat{\nu}}^{(1)}(x\ll1)|^2\approx{2e^{\pi\hat{\nu}}\over\pi\hat{\nu}}$ for $\hat{\nu}\gg1$.
%
Thus the power spectrum of $\delta\phi$ on
superhorizon scales is given as \be P_{\delta\phi}={k^3\over
2\pi^2}|\delta\phi|^2\approx \lf({H\over2\pi}\rt)^2\lf({H\over m_{\phi}}\rt)\lf({k\over
aH}\rt)^3, \ee which is blue-tilted ($\sim k^3$)
with a strongly suppressed factor ${H\over m_{\phi}}$. Hence, the
isocurvature perturbation from $\phi$ is negligible on large
scale.

\begin{figure}[htbp]
\includegraphics[scale=2,width=0.55\textwidth]{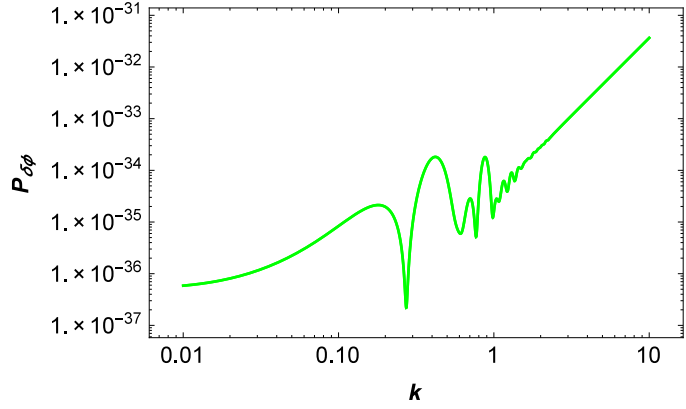}
\caption{The power spectrum of $\delta\phi$ for $\epsilon=0.003$, $A_H=2.72\times10^{-5}$, $c_1=1/8$,
$\alpha'M_p^2=9.6\times10^6$, $V_0/M_p^2=6.5\times 10^{-5}$,
$B=10^{10}$, $\phi_{*}=-2.6$, $\alpha_{initial}=-13.4$ and $\phi_{initial}=-1.5$.} \label{figdeltaphi}
\end{figure}

However, for the quartic potential we used, $\nu$ varies with
time, Eq.(\ref{vv}) isn't able to be solved analytically. But as
long as $V_{\phi\phi}\gg H^2$, $P_{\delta\phi}$ is strongly
suppressed on the large scale. To demonstrate this speculation, we
numerically solve Eq.(\ref{vv}) and plot $P_{\delta\phi}$ in
Fig.\ref{figdeltaphi} with the same parameters in Fig.\ref{fig01}. We
can see that the isocurvature perturbation from $\phi$ is
negligible on large scale. Thus even if after inflation it can be
converted to the curvature perturbation, its effect to the
curvature perturbation from inflaton can be also safely neglected
on the corresponding scale.

\section{The 3+1 decomposition of $R_{GB}^2$}

To derive the perturbation action (\ref{action2}), we
should do the $3+1$ decomposition of the Gauss-Bonnet term
$R_{GB}^2$ \footnote{This result was first obtained by Kaixi Feng. For more details, see Feng's upcoming paper.}.

We follow the ADM metric
\begin{eqnarray}
 g_{\mu\nu}=\left(\begin{array}{cc}N_kN^k-N^2&N_j\\N_i&\gamma_{ij}\end{array}\right),~
 g^{\mu\nu}=\left(\begin{array}{cc}-N^{-2}&\frac{N^j}{N^2}\\ \frac{N^i}{N^2}&\gamma^{ij}-\frac{N^iN^j}{N^2}\end{array}\right)
\end{eqnarray}
where $N$ is the lapse function, $N_i$ is the shift vector and
$\gamma_{ij}$ is the spacial metric. It is useful to define the
normal vector of 3-dimensional hypersurface
$n_\mu=n_0{dt/dx^\mu}=(n_0,0,0)$ and $n^{\mu}=g^{\mu\nu}n_\nu$.
Using the normalization $n_\mu n^\mu=-1$, one have $n_0=N$, which
implies $n_\mu=(N,0,0,0), n^\mu=(-\frac{1}{N},\frac{N^i}{N})$, and
the 3-dimensional induced metric, which is orthogonal to the
normal vector, i.e. $H_{\mu\nu}n^\nu=0$, can be defined to be
$H_{\mu\nu}=g_{\mu\nu}+n_\mu n_\nu$,
\begin{eqnarray}
 H_{\mu\nu}=\left(\begin{array}{cc}N_kN^k&N_j\\N_i&\gamma_{ij}\end{array}\right),~
 H^{\mu\nu}=\left(\begin{array}{cc}0&0\\ 0&\gamma^{ij} \end{array}\right).
\end{eqnarray}
Thus using the 3-dimensional variables, $R_{\mu\nu\rho\sigma}$,
$R_{\nu\sigma}$ and $R$ can be written as
\begin{eqnarray}
R_{\mu\nu\rho\sigma}
&=&~^{(3)}R_{\mu\nu\rho\sigma}+K_{\mu\rho}K_{\nu\sigma}-K_{\mu\sigma}K_{\nu\rho}\nonumber\\
 &~&-(D_\mu K_{\nu\rho}-D_\nu K_{\mu\rho})n_\sigma
    +(D_\mu K_{\nu\sigma}-D_\nu K_{\mu\sigma})n_\rho-(D_\rho K_{\sigma\mu}-D_\sigma K_{\rho\mu})n_\nu\nonumber\\
 & & +(D_\rho K_{\sigma\nu}-D_\sigma K_{\rho\nu})n_\mu-(K_{\alpha\nu}K^\alpha_\rho+D_\rho a_\nu-{\cal L}_nK_{\nu\rho}+a_\nu a_\rho)n_\mu n_\sigma\nonumber\\
 &~& +(K_{\alpha\mu}K^\alpha_\rho+D_\rho a_\mu-{\cal L}_nK_{\mu\rho}+a_\mu a_\rho)n_\nu n_\sigma
 +(K_{\alpha\nu}K^\alpha_\sigma+D_\sigma a_\nu-{\cal L}_nK_{\nu\sigma}+a_\nu a_\sigma)n_\mu
 n_\rho\nonumber\\
    & & -(K_{\alpha\mu}K^\alpha_\sigma+D_\sigma a_\mu-{\cal L}_nK_{\mu\sigma}+a_\mu a_\sigma)n_\nu n_\rho,\\
R_{\nu\sigma}&=&~^{(3)}R_{\nu\sigma}+KK_{\nu\sigma}-2K_{\alpha\sigma}K_\nu^\alpha
   -(D_\alpha K_\nu^\alpha-D_\nu K)n_\sigma-(D_\rho K^\rho_\sigma-D_\sigma K)n_\nu\nonumber\\
 &~&+(K_{\beta\alpha}K^{\alpha\beta}+D_\rho a^\rho-H^{\mu\rho}{\cal L}_n K_{\mu\rho}+a^\rho a_\rho)n_\nu n_\sigma
    -(D_\sigma a_\nu-{\cal L}_n K_{\nu\sigma}+a_\nu a_\sigma),\\
R&=&~^{(3)}R+K^2-3K_{\alpha\sigma}K^{\alpha\sigma}-2(D_\rho
a^\rho-H^{\mu\rho}{\cal L}_n K_{\mu\rho}+a^\rho a_\rho),
\end{eqnarray}
where $D_\mu$ is the covariant derivative with respect to
$H_{\mu\nu}$, ${\cal L}_n$ is the Lie derivative in
$n^\mu$-direction, $K_{\mu\nu}={\cal L}_nH_{\mu\nu}/2$ is the
extrinsic curvature, and $a_\mu\equiv n^\nu\nabla_\nu n_\mu$. Thus
after defining $G_{\mu\nu\rho}=D_\mu K_{\nu\rho}-D_\nu
K_{\mu\rho},~{\cal Z}_{\mu\nu}=D_{(\nu}a_{\mu)}-{\cal L}_n
K_{\mu\nu}+a_\mu a_\nu,~{\cal Z}=g^{\mu\nu}{\cal Z}_{\mu\nu}$, we
have
\begin{eqnarray}
 R_{GB}^2&=&R^2-4R_{\mu\nu}R^{\mu\nu}+R_{\mu\nu\rho\sigma}R^{\mu\nu\rho\sigma}\nonumber\\
 &=&^{(3)}R_{\mu\nu\rho\sigma}~^{(3)}R^{\mu\nu\rho\sigma}+^{(3)}R^2-4~^{(3)}R_{\mu\nu}~^{(3)}R^{\mu\nu}\nonumber\\
 & & +K^4+7(K_{\mu\nu}K^{\mu\nu})^2
   -6~^{(3)}RK_{\mu\nu}K^{\mu\nu}+2^{(3)}R~K^2+16~^{(3)}R_{\mu\nu}K^{\mu\alpha}K^\nu_\alpha\nonumber\\
 &~&+4~^{(3)}R_{\mu\nu\rho\sigma}K^{\mu\rho}K^{\nu\sigma}-14K_{\mu\nu}K^{\nu\rho}K_{\rho\sigma}K^{\sigma\mu}
    +16KK_{\mu\nu}K^{\nu\rho}K_\rho^\mu-10K^2K_{\mu\nu}K^{\mu\nu}\nonumber\\
 &~&-8K~^{(3)}R_{\mu\nu}K^{\mu\nu}-8^{(3)}R_{\mu\nu}n^\mu n^\nu K_{\alpha\beta}K^{\alpha\beta}
    +8R_{\mu\nu\rho\sigma}n^\nu n^\sigma K^{\mu\alpha}K_\alpha^\rho+8G_{\alpha\mu}^{~~~\alpha}G_\beta^{~~\mu\beta}-4G_{\mu\nu\rho}G^{\mu\nu\rho}\nonumber\\
 &~&+16~^{(3)}R_{\mu\nu}n^\mu G_\alpha^{~~\nu\alpha}-8~^{(3)}R_{\mu\nu\rho\sigma}G^{\mu\nu\rho}n^\sigma+8~^{(3)}R_{\mu\nu\rho\sigma}n^\mu n^\rho {\cal Z}^{\nu\sigma}
    -8~^{(3)}R_{\mu\nu}n^\mu n^\nu {\cal Z}+8~^{(3)}R_{\mu\nu}{\cal Z}^{\mu\nu}\nonumber\\
 &~&-4~^{(3)}R {\cal Z}+4{\cal Z}K_{\mu\nu}K^{\mu\nu}+8K{\cal Z}_{\mu\nu}K^{\mu\nu}-8{\cal Z}_{\mu\nu}K^{\mu\alpha}K_\alpha^\nu-4K^2{\cal Z}
\end{eqnarray}
Till now, $R_{GB}^2$ has been decomposed and presented with only
$N$, $N_i$ and $\gamma_{ij}$-related variables. Then, after tedious derivation, the action
(\ref{action2}) of the tensor perturbation can be obtained by
taking $\gamma_{ij}=a^2(t)[e^{h}]_{ij}$, where $h_{ij}$ satisfies
$\partial _i h_{ij}=0$ and $h_{ii}=0$.

\section{The evolution of $\phi(\alpha)$ with different initial value $\phi_{initial}$}

\begin{figure}[htbp]
\includegraphics[scale=2,width=0.55\textwidth]{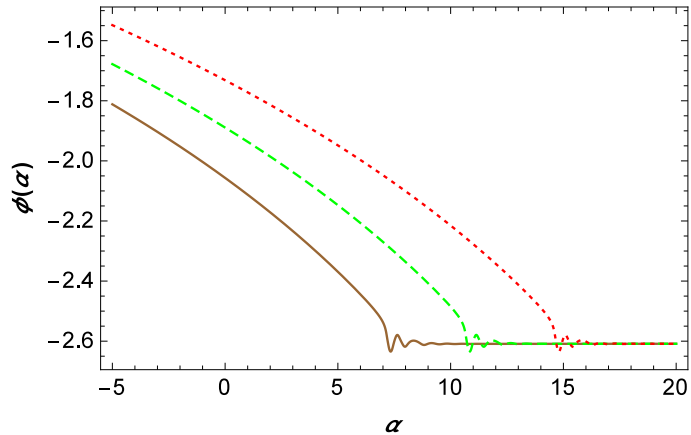}
\caption{The evolution of $\phi(\alpha)$ with different initial value $\phi_{initial}$, where $\phi_{initial}=-1.3$ for red dotted curve, $\phi_{initial}=-1.4$ for green dashed curve, and $\phi_{initial}=-1.5$ for brown solid curve, respectively. We have set $\alpha_{initial}=-13.4$ for these three cases.}\label{diff-phi}
\end{figure}


\end{document}